\input harvmac
\font\symbol=msam10
\font\ensX=msbm10
\font\ensVII=msbm7
\font\ensV=msbm5
\newfam\math
\textfont\math=\ensX \scriptfont\math=\ensVII \scriptscriptfont\math=\ensV
\def\ensemble{\fam\math\ensX}
\overfullrule=0pt
%%%%%%%%%%%%%%%%%%%%%%%%%%%%%%%%%%%%%%%%%%%%%%%%%%%%%%%%%%%%%%%%%%%%%%%%%%%%
%\draftmode
%%%%%%%%%%%%%%%%%%%%%%%%%%%%%%%%%%%%%%%%%%%%%%%%%%%%%%%%%%%%%%%%%%%%%%%%%%%%%
\def\ZZ{{\ensemble Z}}
\def\square{\hbox{\symbol\char'003}}
\def\Str{{\rm Str}}
%%%%%%%%%%%%%%%%%%%%%%%%%%%%%%%%%%%%%%%%%%%%%%%%%%%%%%%%%%%%%%%%%%%%%%%%%%%%%%
\lref\rfDVV{R. Dijkgraaf, E. Verlinde and H. Verlinde, {\sl Matrix
  String Theory},  Nucl. Phys. {\bf B500} (1997) 43, hep-th/9703030.
%%CITATION = HEP-TH 9703030;%%
}
\lref\rfMatrixString{T. Wynter, {\sl Gauge Fields and Interactions in Matrix
String Theory}, Phys. Lett. {\bf B415} (1997) 349, hep-th/9709029.
%%CITATION = HEP-TH 9709029;%%
}
\lref\LukasJK{A.~Lukas and B.~A.~Ovrut, {\sl U-duality symmetries from
the membrane world-volume}, Nucl. Phys. {\bf B502} (1997) 191,
hep-th/9704178. 
%%CITATION = HEP-TH 9704178;%%
}
\lref\deWitVB{B.~de Wit, U.~Marquard and H.~Nicolai,
{\sl Area Preserving Diffeomorphisms And Supermembrane Lorentz Invariance},
Commun. Math. Phys.  {\bf 128} (1990) 39.
%%CITATION = CMPHA,128,39;%%
}
\lref\rfdeWitUnstable{B. de Wit, M. Luscher and H. Nicolai, {\sl The
    Supermembrane is Unstable}, Nucl. Phys. {\bf B320} (1989) 135.
%%CITATION = NUPHA,B320,135;%%
}
\lref\rfKostovVanhove{I.K. Kostov and P. Vanhove, {\sl Matrix string partition
functions}, Phys. Lett. {\bf  B444} (1998) 196, hep-th/9809130.
%%CITATION = PHLTA,B444,196;%%
}
\lref\rfSugino{F. Sugino, {\sl Cohomological field theory approach to matrix
strings},  Int. J. Mod. Phys. {\bf A14} (1999) 3979, hep-th/9904122.
%%CITATION = IMPAE,A14,3979;%%
}
\lref\rfPiolineMembrane{B. Pioline, H. Nicolai, J. Plefka and A. Waldron, 
{\sl $R^4$ couplings, the fundamental membrane and exceptional theta
correspondences},  JHEP {\bf 03} (2001) 036, hep-th/0102123.
%%CITATION = JHEPA,0103,036;%%
}
\lref\rfdeWitPeetersPlefka{B. de Wit, K. Peeters and J.C. Plefka, 
{\sl The Supermembrane with winding}, 
Nucl. Phys. Proc. Suppl. {\bf 62} (1998) 405, hep-th/9707261\semi
%%CITATION = NUPHZ,62,405;%%
B. de Wit, K. Peeters and J. Plefka, {\sl Supermembranes with
winding}, Phys. Lett.  {\bf B409} (1997) 117, hep-th/9705225. 
%%CITATION = PHLTA,B409,117;%%
}
\lref\rfMooreNS{G. Moore, N. Nekrasov and S. Shatashvili, 
{\sl D-particle bound states and generalized instantons}, 
Commun. Math. Phys. {\bf 209} (2000) 77, hep-th/9803265.
%%CITATION = CMPHA,209,77;%%"
}
\lref\rfRusso{J. G. Russo, {\sl Supermembrane dynamics from multiple
interacting strings}, Nucl. Phys. {\bf B492} (1997) 205, hep-th/9610018.
%%CITATION = NUPHA,B492,205;%%"
}
\lref\rfCremmerJuliaScherk{E. Cremmer, B. Julia and J. Scherk, 
{\sl Supergravity Theory in Eleven-Dimensions},  Phys. Lett. {\bf B76} (1978)
409.
%%CITATION = PHLTA,B76,409;%%
}
\lref\rfBergshoeffST{E. Bergshoeff, E. Sezgin and P.K. Townsend, 
{\sl Supermembranes and Eleven-Dimensional Supergravity}, 
Phys. Lett. {\bf 189B} (1987) 75.
%%CITATION = PHLTA,B189,75;%%
}
\lref\rfTownsendMLectures{P.K. Townsend, {\sl Four Lectures on M-theory},
hep-th/9612121.
%%CITATION = HEP-TH 9612121;%%
}
\lref\rfTownsendMtheory{P.K. Townsend, 
{\sl The Eleven-Dimensional Supermembrane Revisited}, 
Phys. Lett. {\bf B350} (1995) 184, hep-th/9501068\semi
%%CITATION = PHLTA,B350,184;%%
E. Witten, {\sl String Theory Dynamics in Various Dimensions}, 
Nucl. Phys. {\bf B443} (1995) 85, hep-th/9503124.
%%CITATION = NUPHA,B443,85;%%
}
\lref\rfBFKOV{
E. Kiritsis and B. Pioline, {\sl On
$R^4$ threshold corrections in IIB string theory and (p,q) string
instantons}, Nucl. Phys. {\bf B508} (1997) 509, hep-th/9707018\semi
%%CITATION = NUPHA,B508,509;%%
C. Bachas, C. Fabre, E. Kiritsis, N.A. Obers and P. Vanhove, 
{\sl Heterotic / type I duality and D-brane instantons}, 
Nucl. Phys. {\bf B509} (1998) 33, hep-th/9707126.
%%CITATION = NUPHA,B509,33;%%
}
\lref\rfGreenGutperle{M.B. Green and M. Gutperle, 
{\sl Effects of D-instantons}, Nucl. Phys. {\bf B498} (1997) 195,
hep-th/9701093. 
%%CITATION = NUPHA,B498,195;%%
}
\lref\rfPiolineObersRpt{B. Pioline and N.A. Obers, 
{\sl U-duality and M-theory}, Phys. Rept. {\bf 318} (1999) 113,
hep-th/9809039.
%%CITATION = PRPLC,318,113;%%
}
\lref\rfPiolineObersI{N.A. Obers and B. Pioline, {\sl Eisenstein series and
string thresholds}, Commun. Math. Phys. {\bf 209} (2000) 275, hep-th/9903113.
%%CITATION = CMPHA,209,275;%%
}
\lref\rfPiolineObersII{N.A. Obers and B. Pioline, 
{\sl Eisenstein series in string theory}, 
Class. Quant. Grav. {\bf 17} (2000) 1215, hep-th/9910115.
%%CITATION = CQGRD,17,1215;%%
}
\lref\rfAusting{P. Austing, {\sl The cohomological supercharge}, 
JHEP {\bf 01} (2001) 009,hep-th/0011211.
%%CITATION = JHEPA,0101,009;%%
}
\lref\rfWittenIndex{E. Witten, {\sl Constraints on Supersymmetry Breaking},
 Nucl. Phys. {\bf B202} (1982)  253.
%%CITATION = NUPHA,B202,253;%%
}
\lref\rfAffleckHW{I. Affleck, J.A. Harvey and E. Witten,
{\sl Instantons And (Super)Symmetry Breaking In (2+1)-Dimensions},
Nucl.  Phys. {\bf B206} (1982) 413.
%%CITATION = NUPHA,B206,413;%%
}
\lref\rfGanorRT{O.J. Ganor, S. Ramgoolam and W. Taylor, 
{\sl Branes, fluxes and duality in M(atrix)-theory}, 
Nucl. Phys. {\bf B492} (1997) 191, hep-th/9611202.
%%CITATION = NUPHA,B492,191;%%
}
\lref\rfVafaWitten{C. Vafa and E. Witten, {\sl A Strong coupling test
of S duality}, Nucl. Phys. {\bf B431} (1994) 3, hep-th/9408074.
%%CITATION = NUPHA,B431,3;%%
}
\lref\DoreyYM{
N.~Dorey, T.~J.~Hollowood and V.~V.~Khoze, {\sl Notes on soliton bound-state
problems in gauge theory and string  theory}, hep-th/0105090.
%%CITATION = HEP-TH 0105090;%%
}
\lref\rfdeWitHoppeNicolai{
B. de Wit, J. Hoppe and H. Nicolai, {\sl On
the quantum mechanics of supermembranes}, Nucl. Phys. {\bf B305}
(1988) 545.
%%CITATION = NUPHA,B305,545;%%
}
\lref\rfTaylor{W. Taylor, {\sl D-brane field theory on compact spaces},
Phys. Lett. {\bf B394} (1997) 282, hep-th/9611042.
%%CITATION = PHLTA,B394,283;%%
}
\lref\rfYi{P. Yi,  {\sl Witten index and threshold bound states of D-branes},
Nucl. Phys. {\bf B505} (1997) 307, hep-th/9704098.
%%CITATION = NUPHA,B505,307;%%
}
\lref\rfSethiStern{
S. Sethi and M. Stern, {\sl D-brane Bound States Redux},
Commun. Math. Phys. {\bf 194} (1998) 675, hep-th/9705046.
%%CITATION = CMPHA,194,675;%%
}
\lref\rfGreenGutperleMatrix{
M.B. Green and M. Gutperle, {\sl D-particle bound states and the D-instanton
measure}, JHEP {\bf 9801} (1998) 005, hep-th/9711107.
%%CITATION = HEP-TH 9711107;%%
}
\lref\rfPorratiR{M. Porrati and A. Rozenberg, {\sl Bound States at
Threshold in Supersymmetrical Quantum Mechanics}, Nucl. Phys. {\bf
B515} (1998) 184, hep-th/9708119.
%%CITATION = NUPHA,B515,184;%%
}
\lref\rfHayakawaI{M.~Hayakawa and N.~Ishibashi,
{\sl Perturbative world-volume dynamics of the bosonic membrane and string},
hep-th/0107103.
%%CITATION = HEP-TH 0107103;%%
}
%%%%%%%%%%%%%%%%%%%%%%%%%%%%%%%%%%%%%%%%%%%%%%%%%%%%%%%%%%%%%%%%%%%%%%%%%%%%%%
\Title{\vbox{\baselineskip12pt
\hbox{hep-th/0107145}
\hbox{SACLAY-SPHT-T01/68}
}}
{U-duality from Matrix Membrane Partition Function}
\smallskip
\centerline{{\bf Fumihiko Sugino and  Pierre Vanhove}}
\centerline{\sl Service de Physique Th{\'e}orique, CEA-Saclay,}
\centerline{\sl F-91191 Gif-sur-Yvette Cedex, France.}
\centerline{\tt sugino, vanhove@spht.saclay.cea.fr}

\bigskip 

We analyse supermembrane instantons (fully wrapped supermembranes) 
by computing the partition
function of the three-dimensional supersymmetrical U(N) matrix
model under periodic boundary conditions. 
By mapping the model to a cohomological field theory and 
considering a mass-deformation of the model,
we show that the partition function exactly leads to the 
U-duality invariant measure factor
entering supermembrane instanton sums. 
On the other hand, a computation based on the quasi-classical assumption 
gives the non U-duality invariant result of the zero-mode analysis 
by Pioline et al. \refs{\rfPiolineMembrane}. 
This is suggestive of the importance of supermembrane self-interactions 
and shows a crucial difference from the matrix string case. 

\Date{July 2001}

%%%%%%%%%%%%%%%%%%%%%%%%%%%%%%%%%%%%%%%%%%%%%%%%%%%%%%%%%%%%%%%%%%%%%%%%%%%
\newsec{Introduction}

The supermembrane (M2-brane) is a mysterious
quantum object. The poor understanding of its three-dimensional
world-volume theory makes difficult to consider it as a fundamental
object. The classical relation between the Cremmer-Julia-Scherk
supergravity \refs{\rfCremmerJuliaScherk} and the background fields
for the supermembrane world-volume theory \refs{\rfBergshoeffST} is
suggestive of the relation between a fundamental object and its
low-energy effective action.  In this case the only dimensionful 
parameter is the eleven-dimensional Planck length $\ell_P$. On the other hand, 
compactified supermembranes give rise to D-branes 
\refs{\rfTownsendMLectures} therefore to non-perturbative string
theory effects.  In dimensions lower than eleven a new parameter (the
eleventh radius $R_{11}$ interpreted as the string coupling constant
\refs{\rfTownsendMtheory}) allows one to make the difference between
fundamental quantum excitations (the fundamental strings) and
solitonic configurations (the D-branes).

 In this setup semi-classical rules for classifying D-brane 
configurations can be derived \refs{\rfGreenGutperle,\rfBFKOV}. The
configurations are characterized by U-duality invariant
number-theoretic functions \refs{\rfPiolineObersRpt}, associated with
the bulk contribution of the Witten index 
for the effective U(N) supersymmetric matrix model description
\refs{\rfYi,\rfSethiStern,\rfGreenGutperleMatrix,\DoreyYM}.
These matrix models  for the collective dynamics of $N$ D-branes are
obtained by compactifying the U(N) supersymmetric quantum
mechanics \refs{\rfTaylor}, originating from a SU(N) regularization
of the light-cone Hamiltonian \refs{\rfdeWitHoppeNicolai} for the
supermembrane
\eqn\eHamiltonian{
{\cal H}= {1\over P_0^+} \int d^2\sigma \sqrt{w} \left[ {P^a P_a\over
2w} + {1\over 4} \{X^a,X^b\}^2 - P_0^+ \bar\theta
\gamma_-\gamma_a\{X^a, \theta\}\right]\ .
}

The purpose of this paper is to exhibit the origin of the U-duality group 
in M-theory compactified on a three-torus  
\eqn\eUduality{
G_U= Sl(3,\ZZ) \times Sl(2, \ZZ)\, , }
by considering Euclidean supermembranes wrapped over the three-torus. 
The $Sl(3,\ZZ)$ group corresponds to the area
preserving group of isometries of the three-torus and $Sl(2,\ZZ)$ acts
on the complex parameter $\Omega= C_{123} + i{\rm Vol}_3$ made from
the v.e.v. of the three-form potential 
and the volume of the three-torus the supermembrane is fully wrapped on.  
The origin of the two groups in~\eUduality\ is understood 
from the matrix model setup \refs{\rfGanorRT}, as the geometrical
$Sl(3,\ZZ)$ symmetry group of reparametrizations of the three-torus,
and a quantum $Sl(2,\ZZ)$ symmetry group exchanging the different
saddle-point contributions of the path integral.

Counting multiply-wrapped Euclidean D-strings 
\refs{\rfBFKOV} consists of including all the
supersymmetric maps of the D-string world-sheet onto the space-time
compactification torus, modulo local and global reparametrizations. This rule
corresponds to the classical saddle-point of the path integral of the {\it
free\/} sigma model for the D-string in the background of the Euclidean
two-torus. A result re-obtained in \refs{\rfKostovVanhove} by computing the
partition function (with the zero-modes subtracted) of the two-dimensional
U(N) matrix string model.  The two-dimensional matrix model can be written as 
a cohomological
field theory for which the quasi-classical approximation is exact
\refs{\rfSugino}, and the problem reduces to sum over free singly connected long
strings wrapping the two-torus \refs{\rfKostovVanhove}.

We explain, in section~3, that the three-dimensional matrix model does
not have exact quasi-classics, therefore the free theory reduction is not
enough. Summing only over the classical configurations of the three-dimensional
matrix model seems to be equivalent to the zero-mode approach to the path integral over
supermembrane configurations of Pioline et al.\refs{\rfPiolineMembrane}, where
all the interactions in~\eHamiltonian\ were discarded. This approximation
leads to the geometrical measure factor

\eqn\eMuWrong{
\hat\mu (N)= \sum_{n|N} n \sum_{p|(N/n)} p^2\, ,
}
which counts the ways to map a volume-$N$ three-torus onto an
Euclidean unit-volume target three-torus, modulo local reparametrizations.
Unlike for the D-string case \refs{\rfBFKOV}, this function~\eMuWrong\ is not
invariant under the full U-duality group~\eUduality\ but is invariant under
$Sl(3,\ZZ)$ alone.  The correct counting of configurations of wrapped supermembranes
is given by the number-theoretic function
\refs{\rfPiolineMembrane,\rfPiolineObersII}

\eqn\eMuCorrect{
\mu (N)= \sum_{n|N} n\, .
}
For $N$ a large prime number, $\hat\mu (N) \simeq N \mu (N)$ meaning
that U-duality equivalent configurations were over-counted by the factor $N$ 
and $\mu (N)$ predicts
the number of ground states for the supermembrane. The
departure of $\hat\mu (N)$ from $N\times \mu (N)$ for finite values of
$N$ shows that the problem is slightly more subtle than an
over-counting.

We explain in section~4 that the
difference between these two functions could be traced back to the
presence of the interaction terms in the Hamiltonian~\eHamiltonian.
 By considering
massive deformations of the ${\cal N}_3=8$ supersymmetric
three-dimensional U(N) matrix model into a ${\cal N}_3=2$ supersymmetric
gauge theory in the cohomological field theory approach, 
the correct counting of configurations~\eMuCorrect\ will be derived.

We conclude, in section~5, with comments about a possible path
integral approach to the supermembrane effects.

%%%%%%%%%%%%%%%%%%%%%%%%%%%%%%%%%%%%%%%%%%%%%%%%%%%%%%%%%%%%%%%%%%%%%%%%%%%
\newsec{The Matrix Model Description}

In order to obtain a matrix model description, we start with the system of $N$
D-particles in (Euclidean) type IIA theory compactified on a two-torus ${\cal T}^2$
parametrized by $x^9$ and $x^{10}$.  The D-particles are wrapped on the time
direction $x^{10}$.  Then, we consider the following duality sequences:
$T_9ST_9$, where $S$ and $T_9$ stand, respectively, for the S-duality and the
T-duality operation with respect to the direction $x^9$.  As a result, we have
the system of $N$ fundamental strings in (Euclidean) type IIA theory on ${\cal
T}^2$, where the fundamental strings are fully wrapped on the ${\cal T}^2$.
This is the same argument as in the derivation of matrix string theory
\refs{\rfDVV} except that the direction $x^{10}$ is compactified. 
One then considers the matrix string theory compactified on a further $S^1$. By
application of the argument for the compactification by Taylor
\refs{\rfTaylor}, we obtain the maximally supersymmetric Yang-Mills theory
with the gauge group U(N) in Euclidean three dimensions.  It is composed by three gauge
connections $A_a$ $(a=1,2,3)$, seven adjoint scalars $X^I$ $(I=1,\cdots,7)$
and sixteen real adjoint fermions $\Psi_\alpha$ $(a=1,\cdots, 16)$:

\eqn\eMatrixModel{\eqalign{
S_{\rm 3D}[A,X,\Psi] = {1\over g^2_{\rm 3D}}  \int_{{\cal T}^3}
d^3\sigma\ \Tr\left[{1\over 4} F_{ab}F^{ab} +\right.&{1\over 2}
(D_{a}X^I)^2 + {i\over 2} \Psi^T  \Gamma^a D_a\Psi\cr 
&\left. - {1\over 4} [X^I,X^J]^2 +{1\over2} \Psi^T \Gamma^I [X_I,\Psi]
\right]\, . 
}}
As usual $D_a = \partial_a - i A_a$. The coupling constant $g^2_{\rm
3D}$ has the dimension $({\rm length})^{-1}$. This model has sixteen real
supercharges (${\cal N}_3=8$) and is invariant under the $Sl(3,\ZZ)$
group of reparametrizations of the Euclidean rectangular three-torus
${\cal T}^3$ with lengths $(R_1,R_2,R_3)$.  This group corresponds to
the $Sl(3,\ZZ)$ appearing in the U-duality group~\eUduality. The
volume of this torus will be denoted by ${\cal V}_o=R_1R_2R_3$.

As in the matrix string case \refs{\rfKostovVanhove}, 
the amplitude of supermembranes fully wrapped on the three-torus in 
M-theory corresponds to the partition function of the super Yang-Mills theory:
\eqn\eZ{
Z^{\rm U(N)}_{{\cal N}_3=8}[ {\cal V}_o] = 
\int {[{\cal D}A]\over {\rm Vol}(U(N))} [{\cal D}X] [{\cal D}\Psi]  
\delta(X^{(0)})\delta(\Psi^{(0)}) \, e^{-S_{\rm 3D}[A,X,\Psi]}\, , 
}
where all the fields obey periodic boundary conditions and 
the zero-mode subtractions are defined as
\eqn\eDelta{\eqalign{
\delta(X^{(0)}) \equiv 
\prod_{I=1}^7 \delta\left(\Tr\int_{{\cal T}^3} d^3\sigma 
X_I\over \sqrt{N{\cal V}_o }\right)&,
\quad
\delta(\Psi^{(0)}) \equiv
\prod_{\alpha=1}^{16} \delta\left(\Tr\int_{{\cal T}^3} d^3\sigma 
\Psi_\alpha\over \sqrt{N{\cal V}_o} \right)\, . 
}}
The path integral measures are normalized using the natural metric on the space 
of small deformations $\delta\varphi$, for $\varphi$ meaning the gauge connection $A_a$ 
or the matter field $X^I$ and $\Psi_\alpha$:
\eqn\eNormalisation{
\int [{\cal D} \delta\varphi] \, \exp\left(- {1 \over 2g_{\rm 3D}^2}\int_{{\cal T}^3} d^3\sigma \, 
\Tr(\delta \varphi)^2\right) =1\, .
}

%%%%%%%%%%%%%%%%%%%%%%%%%%%%%%%%%%%%%%%%%%%%%%%%%%%%%%%%%%%%%%%%%%%%%%%%%%%
\newsec{The Quasi-Classical Calculation}

We show that, if we assume that only the variables along the flat directions
of the potential are relevant and contribute to the partition function (as for
the quasi-classical assumption considered in \refs{\rfKostovVanhove}), we
obtain the non U-duality invariant measure $\hat\mu (N)$ given in~\eMuWrong. The
analysis is performed using the method of \refs{\rfKostovVanhove} with all the
modifications needed for the three-dimensional case.

We project on the flat directions\foot{This limit corresponds to
rescaling the fields by a factor of $g_{\rm 3D}$ and sending $g_{\rm
3D}\to\infty$.} 
\eqn\eClassical{\eqalign{
\Tr([A_a,A_b]^2)=0, &\quad \Tr([A_a,X^I]^2)=0, \quad\Tr([X^I,X^J]^2)=0,\cr
 \quad \Tr (\Psi^T \Gamma^I [X_I,\Psi])=0,& \quad \Tr(\Psi^T \Gamma^a
[A_a,\Psi])=0\, .
}}
That results into breaking the U(N) gauge symmetry to ${\rm U(1)}^N$. 
The fields $\Phi=\{ D_a=\partial_a - i A_a, X^I, \Psi^\alpha\}$ 
can be simultaneously diagonalized by a unitary matrix $V(\sigma^1,\sigma^2,\sigma^3)$ 
such that
\eqn\eDiag{
\Phi(\sigma^1,\sigma^2,\sigma^3)=V^{-1}(\sigma^1,\sigma^2,\sigma^3) 
\Phi^D(\sigma^1,\sigma^2,\sigma^3)V(\sigma^1,\sigma^2,\sigma^3)\, ,
}
with $\Phi^D={\rm diag}\{\Phi_1,\cdots,\Phi_N\}$, 
giving rise to the twisted boundary conditions \refs{\rfKostovVanhove}
\eqn\eBC{\eqalign{
\Phi^D(R_1 ,\sigma^2,\sigma^3)=&S^{-1}\Phi^D(0,\sigma^2,\sigma^3) S,
\quad S=V(0,\sigma^2,\sigma^3)V^{-1}(R_1 ,\sigma^2,\sigma^3),\cr
\Phi^D(\sigma^1,R_2,\sigma^3)=&T^{-1}\Phi^D(\sigma^1,0,\sigma^3) T,
\quad T=V(\sigma^1,0,\sigma^3)V^{-1}(\sigma^1,R_2,\sigma^3),\cr
\Phi^D(\sigma^1,\sigma^2,R_3)=&U^{-1}\Phi^D(\sigma^1,\sigma^2,0) U,
\quad U=V(\sigma^1,\sigma^2,0)V^{-1}(\sigma^1,\sigma^2,R_3)\, ,
}}
where the matrices $S$, $T$ and $U$ act as permutation operators on the
eigenvalues  of the fields $\Phi$. For consistency of the boundary conditions,
the matrices must be mutually commuting. Each triplet of permutations
$(S,T,U)$ describes coverings of the three-torus ${\cal T}^3$ with, in
general, several disconnected components. Each component is interpreted as a
fully wrapped supermembrane over the three-torus. Because each component has
sixteen fermionic zero-modes, saturation of the fermionic zero-modes in the
partition function selects singly connected configurations
\refs{\rfKostovVanhove}. They correspond to various states of a single long
supermembrane wrapping $N$ times the three-torus, reducing the model to a free
U(1) matrix  model on a three-torus of extended size $N{\cal V}_o$ with still
sixteen real supercharges (${\cal N}_3=8$). The large torus is characterized
by the matrix  $M=[m_{ij}]_{1\leq i,j\leq 3}$ with the all entries being
integers and $\det M = N$,  {\sl i.e.} it is spaned by the three vectors
$\vec{\omega}_a=(m_{a1}R_1, m_{a2}R_2, m_{a3}R_3)$.   The periodicity of the
large torus leads to
\eqn\eSTUp{
S^{m_{a1}} T^{m_{a2}} U^{m_{a3}}=1\quad \forall a\in\{1,2,3\}. 
} 
Here, all these equations are not independent. Using $Sl(3,\ZZ)$ transformations, they can be reduced to 
\eqn\eSTU{
U^p=T^nU^k=S^mT^jU^l=1, 
} 
with $N=mnp$, $j=0,\cdots,n-1$ and $k,l=0,\cdots,p-1$. Correspondingly, the matrix $M$ 
becomes 
\eqn\eMap{
M = \pmatrix{ m & j & l\cr 0 & n & k \cr 0 & 0 & p}, 
}
which is a representative of classes modulo the left-action of $Sl(3,\ZZ)$.
Also, the three vectors spanning the corresponding torus $\widetilde{{\cal T}}^3$ are 
\eqn\eVector{
\vec{\omega}_1=(mR_1,jR_2,lR_3), \quad
\vec{\omega}_2=(0,nR_2,kR_3), \quad
\vec{\omega}_3=(0,0,pR_3) \,.
}
An explicit solution to equations~\eSTU\ is 
\eqn\solSTU{\eqalign{
&U= P^{mn}_N, \quad T=P^{-mk}_N \pmatrix{P^m_x &   &   \cr  & \ddots &  \cr  &  & P^m_x}, \cr
&S=P^{kj-nl}_N \pmatrix{P^{-j}_x &  &  \cr  & \ddots  &   \cr  &   &  P^{-j}_x} 
\pmatrix{ P_y &  &  \cr  & \ddots &  \cr  &  & P_y}\, , 
}}
where $x={\rm gcd}(mn,mk,kj-nl)$, $y={\rm gcd}(m,j,kj-nl)$ and for any integer $i$ we define 
${\rm gcd}(i,0)=i$.   
$P_u$ represents a $u\times u$ matrix of a cyclic permutation: 
\eqn\ePu{
P_u=\pmatrix{0 & 1 & 0 & \cdots & 0 \cr 0 & 0 & 1 & \cdots & 0 \cr 
\vdots & \vdots & \vdots &  & \vdots \cr 0 & 0 & 0 & \cdots & 1 \cr 1 & 0 & 0 & \cdots & 0}\, .
}
$P^{mn}_N$, $P_N^{-mk}$ and $P^{kj-nl}_N$ define a covering with  $x$
disconnected components, but the mutually commuting  $S$, $T$ and $U$
represent a single component covering.\foot{In the matrix string case, the
solution (26) in ref. \refs{\rfKostovVanhove} should be corrected so that it
represents a single component covering. For the equations $T^n=T^jS^m=1$ with
$N=mn$ and $j=0,\cdots,n-1$, a correct solution such that $S$ and $T$ are
mutually commuting and form a singly connected covering is as follows:
\eqn\solST{
T=P^m_N, \quad S=P^{-j}_N \pmatrix{P_v &  &  \cr  & \ddots &  \cr  &   & P_v} \, , 
} where $v={\rm gcd}(m,j)$. }
The above solution is a representative modulo appropriate permutations acting on the basis of 
$S$, $T$, $U$. The number of these degrees of freedom is counted to be $(N-1)!$ 
as in the matrix string case~\refs{\rfKostovVanhove}.   
  
In the quasi-classical limit, the partition function reduces to 
the sum over the partition functions of U(1) supersymmetric gauge theory with
the zero-modes subtracted defined on the various tori~\eVector: 
\eqn\eZIR{\eqalign{
Z^{\rm U(N)}_{{\cal N}_3=8}[{\cal V}_o]=
&{(N-1)! \over N!}\sum_{M\atop \det M=N}Z_{[M]}^{\rm U(1)}[N{\cal V}_o] , \cr
Z_{[M]}^{\rm U(1)}[N{\cal V}_o] =& 
\int {[{\cal D}A]\over {\rm Vol}(U(1))} [{\cal D}X] [{\cal D}\Psi]  
\delta(X^{(0)})\delta(\Psi^{(0)}) \, e^{-S^{\rm U(1)}}\, ,
}}
where 
\eqn\eQuasiClassical{
S^{\rm U(1)}={1\over g_{3D}^2}\int_{\widetilde{{\cal T}}^3} d^3\tilde{\sigma}\sqrt{g}
\left[{1 \over 4}g^{ab}g^{cd}F_{ac}F_{bd}
+ {1 \over 2}g^{ab}\partial_aX_I\partial_bX_I 
+{i \over 2} \Psi^T\Gamma^a\partial_a\Psi\right] \, .
}
We introduced the coordinates $\tilde{\sigma}^a$ ranging from 0 to
$|\vec{\omega}_a|$  with the constant metric $g_{ab}$ such
that $\int_{\widetilde{\cal T}^3}d^3\tilde{\sigma}\sqrt{g}=N{\cal V}_o$. 
The denominator in the first formula $N!$ comes from the volume of the permutation group $S_N$ 
which is a part of the original gauge group U(N). 
All the fields except nontrivial flux sectors for the gauge field 
enjoy the periodic boundary condition on the torus~\eVector.  
The contributions from the gauge and the matter fields factorize as 
\eqn\eZsum{
 Z^{\rm U(1)}_{[M]}[N{\cal V}_o]= 
 Z^{\rm U(1)-gauge}_{[M]}[N{\cal V}_o] 
\times (2\pi g_{\rm 3D}^2 \det{}' \square)^{{8\over 2}- {7\over 2}}\, 
} 
where the last factor comes from the integration over the sixteen fermions
and the seven scalars. The prime means the omission of 
zero-modes of the Laplacian $\square \equiv  -g^{ab}\partial_a\partial_b$.  
The first factor is the partition function for the U(1) gauge theory
part, which we now evaluate.

At first, we consider the flux sectors for the gauge field. 
In order to do so, it is convenient to return to the description of the 
${\rm U(1)}^N$-theory on the original torus.  
The fluxes on the original torus are quantized by the first Chern numbers as  
\eqn\eFlux{
\int d\sigma^a d\sigma^b \,\Tr F^D_{ab} = 2\pi n_{ab}, \quad n_{ab}=-n_{ba}\in \ZZ \, ,
}
where the superscript $D$ means a diagonal matrix. 
Thus, we can rewrite the ${\rm U(1)}^N$ gauge field as 
$
A^D_a=-\sum_{b(<a)}f_{ab}\sigma^b {\bf 1}_N+\tilde{A}^D_a, 
$
where $f_{ab}=2\pi{n_{ab} \over NR_aR_b}$ represents a constant magnetic
flux and $\tilde{A}^D_a$ does not generate fluxes globally. 
The value of the classical action for the flux is easily evaluated as 
\eqn\eSflux{
S^{\rm flux}={(2\pi)^2 \over 2Ng^2_{\rm 3D}{\cal V}_o}[q_1^2R_1^2+q_2^2R_2^2+q_3^2R_3^2], 
}
where $q_a$ is an integer dual to $n_{ab}$. 
$\tilde{A}^D_a$ corresponds to a U(1) gauge field on the
extended torus  obeying the periodic boundary condition.

Next we consider 
orthogonal decompositions of the ${\rm U(1)}^N$ gauge field with
respect to the inner product on the space of connections defined by 
\eqn\eInnerproduct{(\delta A^D_{(1)}, \delta A^D_{(2)})\equiv 
\int_{{\cal T}^3} d^3\sigma \, \delta^{ab}\Tr (\delta A^D_{(1)a}
\delta A^D_{(2)b}) \, . 
}
Note that the flux sector is discrete and it does not contribute 
to the continuous variation $\delta A^D$. 
It is easy to see that the following decomposition is possible: 
$\delta A^D_a=\delta\tilde A^D_a= \delta \bar A^D_a+ \delta \hat A^D_a+ 
\partial_a\delta \phi^D$ where
$\hat A^D_a$ is a quantum fluctuation 
satisfying the Lorentz gauge condition $\partial \cdot \hat A^D=0$ and 
$\int_{{\cal T}^3} d^3\sigma\, \Tr \hat A^D_a=0$ and 
$\phi^D$ is a ${\rm U(1)}^N$ gauge function connected to the identity, 
with 
$\int_{{\cal T}^3} d^3\sigma\, \Tr \phi^D=0$.\foot{Since 
the constant zero-modes of $\phi^D$ cause no gauge transformation, we do
not consider.} 
$\bar A^D_a$ is a constant zero-mode (a flat connection), 
which is immediately seen to be proportional to the unit matrix: $\bar A^D_a=c_a {\bf 1}_N$ 
due to the periodic boundary conditions modulo $S_N$ permutations. 
The measure becomes
\eqn\eMeasure{
{[{\cal D}\tilde{A}^D]\over {\rm Vol}(U(1)^N)} =
\left (\prod_{a=1}^3 dc_a\right)\, 
{[{\cal D}\hat A^D]\,[{\cal D}\partial \phi^D]\over {\rm Vol}(U(1)^N)} \, . 
}
Because the gauge zero-mode $c_a$ lives on a circle of  circumference ${2\pi \over R_a}$, 
the integrals over the zero-modes  give the result 
\eqn\eIntca{
\int \prod_{a=1}^3 dc_a = {(2\pi)^3 \over {\cal V}_o}. 
}
Returning to the description of the U(1)-theory on the extended torus, we have 
\eqn\eZUone{\eqalign{
&Z^{\rm U(1)-gauge}_{[M]}[N{\cal V}_o]=\left(\sum_{\rm flux}e^{-S^{\rm flux}}\right) 
{(2\pi)^3 \over {\cal V}_o}  \cr 
&\times \int[{\cal D}\hat A]\,{ [{\cal D} \phi]\over {\rm Vol}(U(1))}
(\det{}' \square)^{{1\over 2}} \, 
\exp\left(-{1\over 4g^2_{\rm 3D}} \int_{\widetilde{\cal T}^3} d^3\tilde\sigma^a\  
\hat{F}_{ab}\hat{F}^{ab}\right)\, , 
}}
where $\hat{F}_{ab}$ is a field strength corresponding to $\hat{A}_a$. 
$\hat{A}_a$ and $\phi$ are the variables of the U(1) gauge theory 
corresponding to $\hat{A}^D$ and $\phi^D$, respectively.  

From the usual definition of the gauge volume ${\rm Vol}(U(1))$ which concerns elements connected 
to the identity and includes the constant modes of the gauge function, 
we have    
\eqn\eDphi{
\int [{\cal D} \phi] = {{\rm Vol}(U(1))\over 2\pi}\, .
}
We dualize the field strength into a vector $\hat{f} =\star \hat{F} =\star d\hat{A}$, 
and consider a change of the variables from $\hat{A}$ to $\hat{f}$.  
The Jacobian for this change is computed by remarking that the inner product 
\eqn\eJacobian{
(\delta\hat{f},\delta\hat{f}) = 
\int_{\widetilde{\cal T}^3} d^3\tilde{\sigma}\sqrt{g}g_{ab}\,\delta\hat{f}^a\delta\hat{f}^b
=\int_{\widetilde{T}^3} d^3\tilde{\sigma}\sqrt{g}\,\delta{\hat{A}}^a\square P_a{}^b\delta{\hat{A}}_b \, 
}
implies $[{\cal D} \hat{f}]=[\det{}'\square P_a{}^b]^{1/2}[{\cal D}
\hat{A}]$ with $P_a{}^b$ being the projection operator into transverse
directions: $P_a{}^b=\delta_a{}^b+{\partial_a\partial^b \over
  \square}$.  Since $P_a{}^b$ has the eigenvalues 0 and 1 with
multiplicity 1 and 2 respectively, we have $[{\cal D}
\hat{f}]=(\det{}'\square)\, [{\cal D} \hat{A}]$.  The last factor of
the partition function~\eZUone\ becomes
\eqn\esecond{
{1\over 2\pi} \int  [{\cal D}\hat f]\,
(\det{}' \square)^{{1\over 2}-1} \, 
\exp\left(-{1\over 2g^2_{\rm 3D}} \int_{\widetilde{\cal T}^3} d^3\tilde\sigma\sqrt{g}  
\hat{f}_{a}\hat{f}^{a}\right)\, , 
}
Noticing that $\hat{f}$ has no zero-modes, from the
normalization~\eNormalisation\ follows
\eqn\ef{
\int [{\cal D}\hat{f}]\ \exp\left(-{1\over 2g^2_{\rm 3D}} \int_{\widetilde{\cal T}^3}
d^3\tilde\sigma\sqrt{g} \ \hat{f}_a \hat{f}^a\right)= (2\pi g_{\rm 3D}^2)^{-3/2}\, .
}

Collecting everything, the determinant factors of the Laplacian are cancelled, 
the partition function~\eZ\ in the quasi-classical approximation becomes 
\eqn\eZapprox{
Z^{\rm U(N)}_{{\cal N}_3=8}[{\cal V}_o] =
{2\pi \over Ng_{\rm 3D}^2{\cal V}_o}\sum_{[M]\atop \det M=N} \, 
\left(\sum_{\rm flux}e^{-S^{\rm flux}}\right)\, .
}
In the M-theory side, the fluxes are interpreted as Kaluza-Klein states (wrapped supergravitons) 
dissolved in the wrapped membrane.  Since the flux sum is independent of the (discretized) 
``moduli'' $M$ of $\widetilde{\cal T}^3$, 
the sum over the ``moduli'' leads to the number theoretic factor~\eMuWrong\ as 
$$
\sum_{[M]\atop \det M=N}  1 =\sum_{mnp=N}np^2
= \sum_{n|N} n\sum_{p|(N/n)} p^2 
= \hat\mu (N), 
$$
which counts the number of wrapping a size-$N$ $\widetilde{\cal T}^3$ over 
a unit-volume ${\cal T}^3$ up to local reparametrizations. 
It should be noticed that the origin of $\hat\mu (N)$ from the sum over the
``moduli''  is same as in the calculation in ref. \refs{\rfPiolineMembrane}. 

Finally we obtain the result of the quasi-classical calculation as 
\eqn\eZclassical{\eqalign{
\left.Z^{\rm U(N)}_{{\cal N}_3=8}[{\cal V}_o]\right|_{\rm quasi-classical} =& 
{2\pi \over g_{\rm 3D}^2{\cal V}_o}{\hat\mu (N)\over N} \cr 
 & \times \sum_{q_i\in\ZZ}\exp\left(-{(2\pi)^2 \over 2Ng^2_{\rm 3D}{\cal V}_o}
[q_1^2R_1^2+q_2^2R_2^2+q_3^2R_3^2]\right)\, . 
}}

We should remark that the result derived from the quasi-classical assumption 
respects the $Sl(3,\ZZ)$ symmetry
of the model but breaks the $Sl(2,\ZZ)$ symmetry, as can be seen on the end
result~\eZclassical.

%%%%%%%%%%%%%%%%%%%%%%%%%%%%%%%%%%%%%%%%%%%%%%%%%%%%%%%%%%%%%%%%%%%%%%%%%%%
\newsec{The Cohomological Field Theory Approach}

In this section we follow the method of \refs{\rfSugino} (for different
approaches to the cohomological matrix models see for instance
\refs{\rfMooreNS,\rfAusting}) that consists in deforming a three-dimensional
cohomological field theory with sixteen real supercharges, which is equivalent
to the matrix membrane model, by adding a mass perturbation such that the
classical saddle point of the deformed theory is a three-dimensional gauge
theory with four real supercharges (${\cal N}_3=2$).\foot{For a  similar
analysis in the context of super Yang-Mills quantum mechanics, see \refs{\rfPorratiR}.}
In the matrix string case (two-dimensional ${\cal N}_2=8$ super Yang-Mills
theory), the results from the two methods coincide, because the theory after
the mass deformation is an ${\cal N}_2=2$ super Yang-Mills theory, which can
again be written as a cohomological field theory~\refs{\rfSugino}, and the
calculation can be completely reduced to a classical saddle point problem.
As we will see, the major difference between the matrix string and the matrix
membrane models is that the mass-deformed matrix membrane model is not
equivalent to a cohomological model any longer~\refs{\rfSugino}, therefore the
computation will not entirely localize on the classical configurations.

The overall U(1) part of the gauge group U(N) decouples and can be treated as
a free field theory. We use the mass deformation method on the remaining
non-Abelian part of the partition function~\eZ\ as in \refs{\rfSugino}:
\eqn\eSep{
Z^{\rm U(N)}_{{\cal N}_3=8}[{\cal V}_o]= Z^{\rm U(1)}_{{\cal N}_3=8}[{\cal V}_o]
\times Z^{\rm SU(N)/\ZZ_N}_{{\cal N}_3=8}[{\cal V}_o].  
}
$Z^{\rm U(1)}$ denotes the partition function of the overall U(1)
of the U(N) theory with the zero-mode delta functions~\eDelta\
inserted. The gauge field is expanded as 
$A_a=A^{\rm U(1)}_a{\bf 1}_N+A^r_aT^r$, 
where $T^r$'s span a basis of SU(N)-generators. The
matter fields are decomposed similarly. First, we consider the partition function of the SU(N) theory 
$Z^{\rm SU(N)}_{{\cal N}_3=8}[{\cal V}_o]$ (in the zero 't Hooft discrete
flux sector). The action is exactly in the form of the dimensional
reduction of four-dimensional ${\cal N}_4=4$ SU(N) supersymmetric
Yang-Mills theory, so the argument below is a dimensionally reduced
version of the four-dimensional theory.  We map the theory to a
cohomological field theory by twisting \refs{\rfVafaWitten} and
calculate the partition function by adding the mass-perturbation
\eqn\eMass{
\Delta S=-{m_{\rm 3D} \over2\sqrt2g_{\rm 3D}^2}\int_{{\cal T}^3} d^3\sigma \
\Tr (\Phi_1^2+\Phi_2^2+\Phi_3^2)|_{\theta\theta}
+ {\rm h.c.}, 
}
where $\Phi_s\, (s=1,2,3)$ represents the chiral superfields 
in the four-dimensional ${\cal N}_4=1$ superfield formalism. 
After integrating out the chiral superfields in the mass-perturbed theory, 
contributions to the path integral localize on
classical vacua described by three-dimensional ${\cal N}_3=2$ 
supersymmetric ${\rm SU(n)}\otimes \ZZ_m$  gauge theory with $N=mn$ \refs{\rfSugino}. 
As a result, the partition function $Z^{\rm SU(N)}_{{\cal N}_3=8}[{\cal V}_o]$ becomes 
\eqn\eZcoho{ 
Z^{\rm SU(N)}_{{\cal N}_3=8}[ {\cal V}_o] =\sum_{mn=N}Z^{\rm SU(n)\otimes \ZZ_m}[ {\cal V}_o]
=  \sum_{mn=N} m^2 Z^{\rm SU(n)}_{{\cal N}_3=2}[ {\cal V}_o]\, ,
}
where the factor $m^2=m^{3-1}$ comes from the summation over the flat $\ZZ_m$-bundles. 

The three-dimensional 
${\cal N}_3=2$ SU(n) gauge theory with the periodic boundary conditions, 
has a single Higgs scalar field with a continuous spectrum 
beginning at zero-energy. Therefore, the relation between 
$Z^{\rm SU(n)}_{{\cal N}_3=2}[ {\cal V}_o]$ and the Witten index is unclear. 
Seeing the three-dimensional gauge model as the dimensional reduction
of the four-dimensional gauge theory, the partition function 
$Z^{\rm SU(n)}_{{\cal N}_3=2}[ {\cal V}_o]$
is identified with the bulk part of the Witten
index for the four-dimensional SU(n) gauge theory with four
real supercharges (${\cal N}_4=1$): 
\eqn\eReduction{
I_W^{{\cal N}_4=1}(SU(n)) \equiv \lim_{\beta\to\infty} \Str\left( e^{-\beta H_{\rm 4D}}\right) 
= \lim_{\beta\to0} \Str\left( e^{-\beta H_{\rm 4D}}\right) +  \delta I_W \, ,
}
where the fields obey the periodic boundary condition with respect to all the
four directions. As will be seen just below, the modes which are constant 
with respect to the time variable dominate the supertrace term of the right hand side, 
in the limit $\beta \to 0$. It will result the three-dimensional model with
the periodic boundary condition we are considering. However, the four-dimensional
gauge theory in the finite box has a discrete spectrum, so the deficit term
$\delta I_W$ vanishes. The first term in r.h.s. can be regarded as the
partition function of the four-dimensional theory,  where the fourth direction
is a tiny circle of length
$\beta$: $\lim_{\beta\to0} Z^{\rm SU(n)}_{{\cal N}_4=1}[ {\cal V}_o\times [0,\beta]]$. 
The action of the ${\cal N}_4=1$ four-dimensional SU(n) gauge model is 
\eqn\eLFourD{
S_{\rm 4D} =  \int_0^\beta dt \int_{{\cal T}^3} d^3\sigma\,
\left[ {1\over g^2_{\rm 4D}} \Tr\left({1\over 4} F_{ab} F^{ab} 
+ {i\over 2} \bar\lambda \Gamma\cdot \partial \lambda   \right) 
\right]\, .
}
In the limit $\beta\to0$ with $g_{\rm 3D}^{-2}=\beta g_{\rm 4D}^{-2}$ kept
fixed, because of the  discreteness of its spectrum 
the theory smoothly flows to the three-dimensional ${\cal N}_3=2$
theory.\foot{This is a similar argument as in  \refs{\rfYi} for the super
Yang-Mills quantum mechanics.  However, our case is more obvious because the
spectrum is discrete.}  On the other hand, the value of the Witten index is
known from ref.~\refs{\rfWittenIndex} to be $I_W^{{\cal N}_4=1}(SU(n)) = n$. 
Therefore, we get   
\eqn\eZperiodic{
Z^{\rm SU(n)}_{{\cal N}_3=2}[ {\cal V}_o]=n \, .
}
Plugging this result into~\eZcoho, we obtain for the ${\rm SU(N)}/\ZZ_N$
partition function with periodic boundary conditions
\eqn\eZSUcoho{
Z^{\rm SU(N)/\ZZ_N}_{{\cal N}_3=8}[{\cal V}_o]=
{1\over N^2}Z^{\rm SU(N)}_{{\cal N}_3=8}[ {\cal V}_o] 
={1\over N^2}  \sum_{mn=N} m^2 n = {1 \over N} \sum_{m|N} m ={1\over N}\ \mu (N), 
}
which is the U-duality invariant function~\eMuCorrect.

 Next, we consider the contribution of the overall U(1)-part $Z^{\rm U(1)}[{\cal V}_o]$. 
The action is 
\eqn\eSoverallUone{
S^{\rm U(1)}={N \over g_{\rm 3D}^2}\int_{{\cal T}^3}d^3\sigma 
\left[{1 \over 4}(F^{\rm U(1)}_{ab})^2+{1 \over 2}(\partial_aX^{\rm U(1)})^2
+{i \over 2}\Psi^{{\rm U(1)}T}\Gamma_a\partial_a\Psi^{\rm U(1)}\right]. 
}
The path integral is performed with the normalizations~ \eNormalisation :
\eqn\eUonenormalisation{
\int [{\cal D} \delta \varphi^{\rm U(1)}]\,\exp\left(-{N \over 2g_{3D}^2}\int_{{\cal T}^3}d^3\sigma\ 
(\delta\varphi^{\rm U(1)})^2\right)=1. 
}
Contributions from the fluxes and the flat connections are evaluated by the same way as in the 
quasi-classical case~\eSflux\ and \eIntca. 
The integrals for the matter part give the factor
$(2\pi g_{\rm 3D}^2 \det{}'\square)^{1/2}$, and the same procedure as
in the quasi-classical computation leads to the result
\eqn\eZoverallUone{
Z^{\rm U(1)}[{\cal V}_o]={2\pi \over g_{\rm 3D}^2{\cal V}_o}
\sum_{q_i\in\ZZ}\exp\left(-{(2\pi)^2 \over 2Ng^2_{\rm 3D}{\cal V}_o}
[q_1^2R_1^2+q_2^2R_2^2+q_3^2R_3^2]\right) \, ,
}
where $q_i$'s represent the fluxes. 

Now we find the final form of the cohomological field theory calculation as 
\eqn\efinalcoho{
Z^{\rm U(N)}_{{\cal N}_3=8}[{\cal V}_o]={2\pi \over g_{\rm 3D}^2{\cal V}_o}\, 
{\mu (N) \over N}\sum_{q_i\in\ZZ}\exp
\left(-{(2\pi)^2 \over 2Ng^2_{\rm 3D}{\cal V}_o}[q_1^2R_1^2+q_2^2R_2^2+q_3^2R_3^2]\right) \, .
}
In the zero-flux sector, the result~\efinalcoho\ reproduces the U-duality
invariant counting~\eMuCorrect\ of wrapped Euclidean supermembrane
configurations over a three-torus. Nonzero-flux sectors show the contribution 
from Kaluza-Klein states of supergravitons dissolved in the wrapped
supermembrane, which is also consistent to the U-duality. It should be
remarked that in eq.~\eZcoho\ the three-dimensional ${\cal N}_3=2$ theory can
not be described as a cohomological field theory by the twisting procedure
because the theory has only a single Higgs field. The twisting needs at least
two Higgs fields. So it is possible that quantum fluctuations contribute
nontrivially to the partition function of the ${\cal N}_3=2$ theory. In other
words, the Higgs field yields a continuous spectrum, and thus it is likely
that the cancelation of the contributions from the Higgs and its superpartner
is not precisely realized, 
which is analogous to the situation of the appearance of the deficit terms
in the Witten index calculation for super Yang-Mills quantum mechanics
\refs{\rfYi,\rfSethiStern}.  As mentioned at the beginning of this section,
this situation forms a sharp contrast with the matrix string
case. It seems to be a reason of the failure of the quasi-classical approach
and at the same time to show a crucial difference between strings and
membranes.

It would be interesting to speculate 
how the result changes when the continuous zero-modes of the Higgs 
and its superpartner are removed. 
Let us consider the ${\cal N}_3=2$ supersymmetric SU(n) gauge theory with the
twisted boundary conditions \refs{\rfSugino}:
\eqn\eBCtwisted{\eqalign{
A_a(\sigma_1,\sigma_2,\sigma_3) = & A_a(\sigma_1+R_1,\sigma_2,\sigma_3)\cr
= & P A_a(\sigma_1,\sigma_2+R_2,\sigma_3)P^{-1}\cr    
= & Q A_a(\sigma_1,\sigma_2,\sigma_3+R_3)Q^{-1}\, ,
}} 
where $P$ and $Q$ are 't Hooft clock and shift matrices satisfying
$PQP^{-1}Q^{-1}\in \ZZ_n$. All the other fields also obey the same boundary
conditions. In this case, the zero-modes of the Higgs field are absent, and
the spectrum is discrete. The partition function is equal to the Witten index,
which is known to be 1 from ref.~\refs{\rfAffleckHW}. Thus the value of the
partition function of the non-Abelian part without the zero-modes of the Higgs
multiplet will become
\eqn\eTwistedBC{
Z^{\rm SU(N)/\ZZ_N}_{{\cal N}_3=8} =
{1 \over N^2}\sum_{N=mn}m^2 \left.Z^{\rm SU(n)}_{{\cal N}_3=2}\right|_{\rm twisted\ b.c.}=
{1\over N^2} \sum_{m|N} m^2\, .
}
This reproduces the measure factor entering D-instanton 
(supergraviton in the context of M-theory) effects
\refs{\rfYi,\rfSethiStern,\rfGreenGutperleMatrix,\rfMooreNS}.

%%%%%%%%%%%%%%%%%%%%%%%%%%%%%%%%%%%%%%%%%%%%%%%%%%%%%%%%%%%%%%%%%%%%%%%%%%%
\newsec{Discussion}

The analysis of the dynamics of wrapped supermembranes has already
been the subject of various papers
\refs{\rfRusso,\rfdeWitPeetersPlefka}, but a complete understanding of
the supermembrane as a fundamental object is still laking (despite an
interesting recent attempt \refs{\rfPiolineMembrane}). The main
difficulties rely on needle-like deformations that cost no energy
\refs{\rfdeWitUnstable}, which are likely to survive after compactifications 
\refs{\rfdeWitPeetersPlefka}. As an additional complication, a naive
generalization of the matrix model regularization \refs{\deWitVB} of
the flat space light-cone supermembrane was shown to fail
\refs{\rfdeWitPeetersPlefka}, because of the difficulty for the
structure constants to satisfy both the Jacobi identity and the
periodicity conditions around the compact directions. Consequently, 
strictly speaking, a direct and rigorous 
derivation of the matrix membrane  model~\eMatrixModel\ from wrapped
supermembranes on a three-torus is not  known. 

We showed in this paper that the membrane matrix model
contains enough information for reproducing the measure factor of
wrapped supermembranes, and exhibiting the full $Sl(3,\ZZ)\times 
Sl(2,\ZZ)$ symmetries of the moduli space of the model. Again a direct
analysis of these symmetries from the perspective of the supermembrane
world-volume action turns to be more subtle than the naive
generalization of string approach \refs{\LukasJK,\rfPiolineMembrane}.

 It is remarkable that turning on/off the continuous zero-modes of the Higgs
multiplet in the ${\cal N}_3=2$ theory lead to respectively the supermembrane
instantons measure factor and D-instanton (supergraviton) measure factor.  The
former corresponds to the system of the bound states of the supermembrane and
supergravitons, and the latter to the system of the supergraviton states
alone. This is again suggestive of the Higgs zero-modes being identified with
an essential ingredient of the supermembrane.

Differently from the matrix string case \refs{\rfDVV}, where by
dimensional analysis one can be convinced of the appearance of a
conformal field theory description in the infra-red limit, we will not be able to 
expect completely analogous phenomena for membranes. 
This is supported by the failure of the
infra-red (strong coupling) limit analysis of section~3 to reproduce
the correct measure factor~\eMuCorrect.  Consequently, reconstructing
the interactions between supermembranes will take a different route
than in \refs{\rfMatrixString}.\foot{For a recent interesting attempt toward 
perturbative formulation for membranes, see ref. \refs{\rfHayakawaI}.}  
Rederiving these results from a direct
supermembrane path integral analysis is an important problem, that is
left for a future publication.

%%%%%%%%%%%%%%%%%%%%%%%%%%%%%%%%%%%%%%%%%%%%%%%%%%%%%%%%%%%%%%%%%%%%%%%%%%%
\bigskip
{\bf Acknowledgements: } We thank Adi Armoni, Shyamoli Chaudhuri, 
Fran{{\c{}}c}ois David, 
Hiroyuki Fuji, Masafumi Fukuma, Nobuyuki Ishibashi, Elias Kiritsis, 
Ivan Kostov, Tsunehide Kuroki, Tadakatsu Sakai and Tamiaki Yoneya 
for very useful discussions, as well as Jan Plefka for email
correspondence. P.V. thanks all the participants of the superstring theory
meeting and the people of the physics department of the university of Crete
at Heraklion for discussions.  Also, F.S. thanks participants and organizers 
of the string theory symposium at Tohwa university, Fukuoka, Japan, and 
members of theory group at KEK for discussions and hospitality during his stay.  
P.V. received partial support from the EEC
contract HPRN-CT-2000-00122.
%%%%%%%%%%%%%%%%%%%%%%%%%%%%%%%%%%%%%%%%%%%%%%%%%%%%%%%%%%%%%%%%%%%%%%%%%%%
\listrefs
\bye